\documentclass[]{spie}  
\pdfoutput=1

\usepackage{amsmath,amsfonts,amssymb}
\usepackage{graphicx}
\usepackage[colorlinks=true, allcolors=blue]{hyperref}
\usepackage{multirow}
\usepackage{subcaption}
\usepackage[table]{xcolor}
\usepackage{booktabs}
\usepackage{amsmath} 
\newcommand{\angstrom}{\textup{\AA}}
\usepackage[table]{xcolor}
\definecolor{lightgray}{gray}{0.9}
\definecolor{darkgray}{gray}{0.75}

\title{Recent X-ray hybrid CMOS detector developments and measurements}

\author[a]{Samuel V. Hull}
\author[a]{Abraham D. Falcone}
\author[a]{David N. Burrows}
\author[a]{Mitchell Wages}
\author[a]{Tanmoy Chattopadhyay}
\author[a]{Maria McQuaide}
\author[a]{Evan Bray}
\author[a]{Matthew Kern}
\affil[a]{Pennsylvania State University, Department of Astronomy and Astrophysics, 525 Davey Lab, University Park, Pennsylvania 16802, United States}

\authorinfo{Further author information: (Send correspondence to S.V.H.)\\S.V.H.: E-mail: s.hull@psu.edu}

\begin{document} 
\maketitle

\begin{abstract}
The Penn State X-ray detector lab, in collaboration with Teledyne Imaging Sensors (TIS), have progressed their efforts to improve soft X-ray Hybrid CMOS detector (HCD) technology on multiple fronts. Having newly acquired a Teledyne cryogenic SIDECAR$^{\text{TM}}$ ASIC for use with HxRG devices, measurements were performed with an H2RG HCD and the cooled SIDECAR$^{\text{TM}}$. We report new energy resolution and read noise measurements, which show a significant improvement over room temperature SIDECAR$^{\text{TM}}$ operation. Further, in order to meet the demands of future high-throughput and high spatial resolution X-ray observatories, detectors with fast readout and small pixel sizes are being developed. We report on characteristics of new X-ray HCDs with 12.5 micron pitch that include in-pixel CDS circuitry and crosstalk-eliminating CTIA amplifiers. In addition, PSU and TIS are developing a new large-scale array Speedster-EXD device. The original $64\times64$ pixel Speedster-EXD prototype used comparators in each pixel to enable event driven readout with order of magnitude higher effective readout rates, which will now be implemented in a $550\times550$ pixel device. Finally, the detector lab is involved in a sounding rocket mission that is slated to fly in 2018 with an off-plane reflection grating array and an H2RG X-ray HCD. We report on the planned detector configuration for this mission, which will increase the NASA technology readiness level of X-ray HCDs to TRL 9.
\end{abstract}


\section{INTRODUCTION}
\label{sec:intro}  

Looking beyond \textit{Chandra}, concepts for future large X-ray astrophysics missions have already been well developed \cite{smartx}. The \textit{Lynx} (previously X-ray Surveyor) mission concept is the current plan to be put forward by the U.S. X-ray community for NASA's 2020 Decadal Survey \cite{XRS}. This mission calls for over 30 times the collecting area of \textit{Chandra}, which will enable \textit{Lynx} to tackle science goals not accessible to the current generation of high energy observatories by opening up our window into the high-redshift and low luminosity universe. The high throughput needs of future X-ray missions require focal plane detectors with faster readout speed capabilities than modern X-ray Charge-Coupled Devices (CCDs), which would be subject to substantial saturation effects\cite{pileup}. Hybrid CMOS detectors (HCDs) are active pixel sensors that offer the fast readout speeds needed for future missions like \textit{Lynx}, while also possessing greater radiation hardness and lower power requirements . 

In these proceedings we present new measurements of several types of hybrid CMOS devices, including read noise and energy resolution measurements of an H2RG detector with a Teledyne cryogenic SIDECAR$^{\text{TM}}$ ASIC \cite{sidecar}, and the first measurements with small pixel (12.5 $\mu$m) HCDs. These small pixel HCDs, which are uniquely suited for high angular resolution missions like \textit{Lynx}, contain in-pixel correlated double sampling (CDS) subtraction and a CTIA amplifier that eliminates interpixel capacitance. In addition, we report on other developments and future plans for X-ray HCDs, which include the launching of an H2RG detector on a sounding rocket in the Spring of 2018 and plans for a large format detector with sparse readout capabilities.

\section{HYBRID CMOS DETECTORS}
\label{sec:hcd}

\begin{figure} [h]
   \begin{center}
    \begin{minipage}[b]{0.45\textwidth}
    \includegraphics[width=\textwidth]{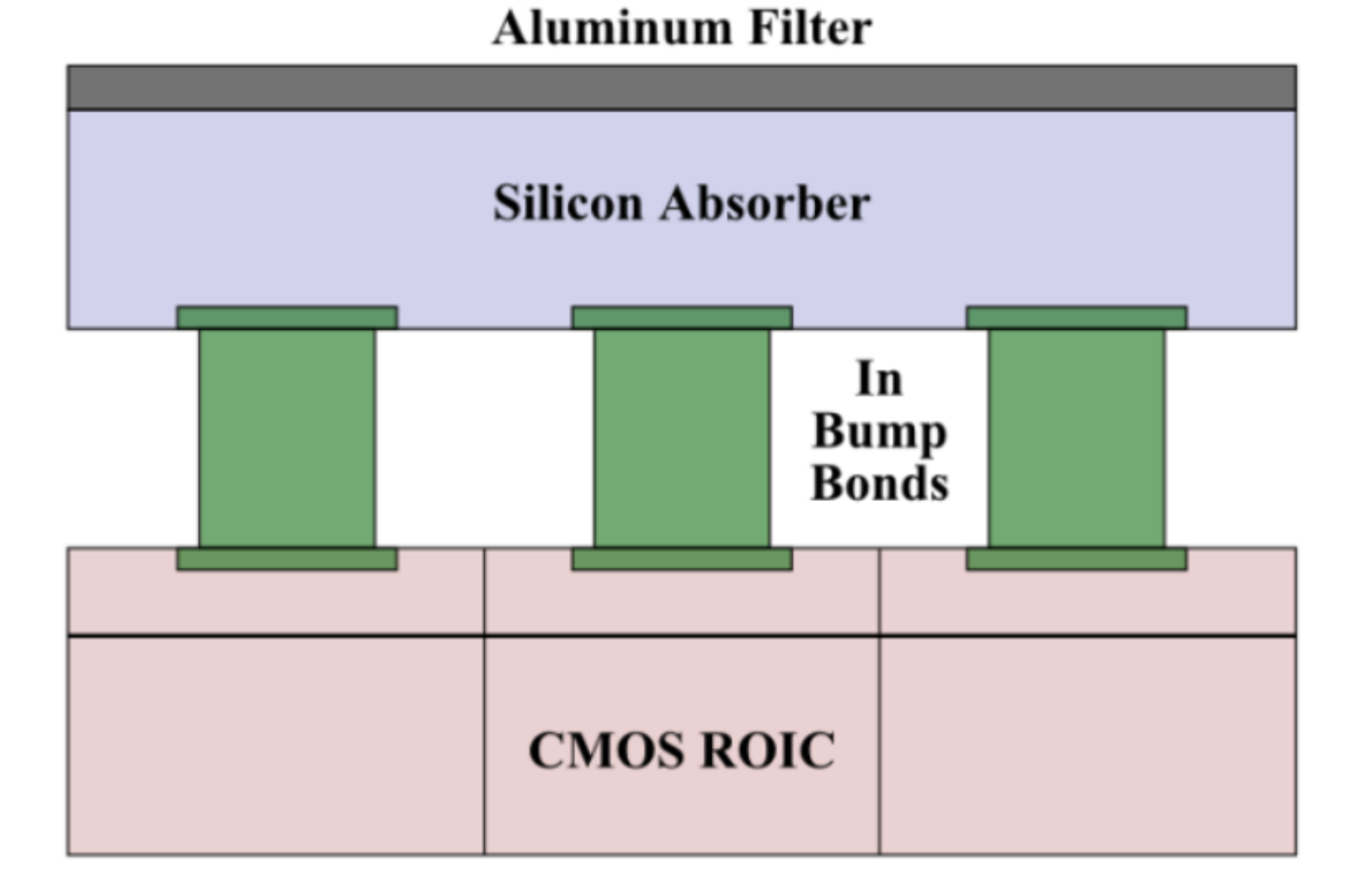}
    \caption{\label{fig:hcd} Cross-sectional schematic of hybrid CMOS X-ray detector. Si absorbing layer is bump bonded to the read out electronics layer in each pixel, allowing for separate optimization.}
  \end{minipage}
  \hfill
  \begin{minipage}[b]{0.45\textwidth}
    \includegraphics[width=\textwidth]{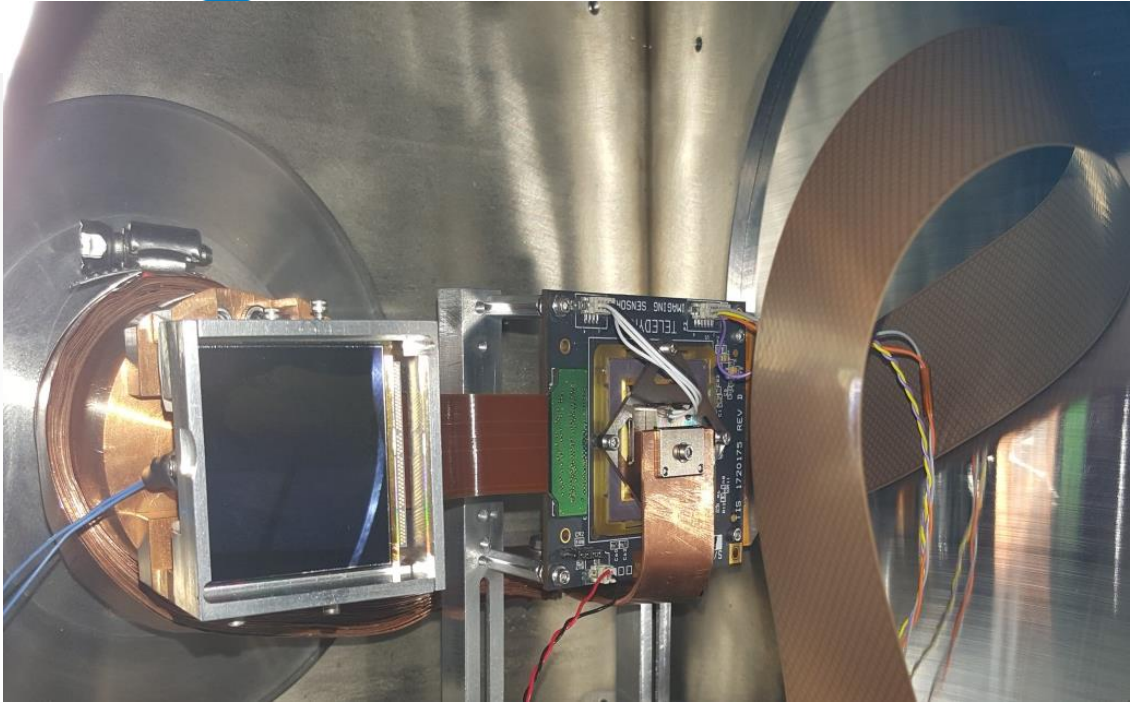}
    \caption{\label{fig:cube} The interior of our test chamber with an H2RG hybrid CMOS detector attached to cold finger and the cryogenic SIDECAR$^{\text{TM}}$ to the right of the detector.}
  \end{minipage}
   \end{center}
   \end{figure}

X-ray hybrid CMOS detectors are a type of X-ray-sensitive active pixel sensor being developed as a joint collaboration between Penn State University (PSU) and Teledyne Imaging Sensors (TIS), and are based on the TIS HAWAII hybrid Si CMOS detectors \cite{hawaii}. While possessing similarities to traditional CCDs, HCDs are a different detector technology. X-ray HCDs are composed of two separate layers: the absorbing silicon layer and the readout integrated circuit (ROIC). The absorbing layer is responsible for photon-to-charge conversion via photoelectric absorption in the silicon pixel array, while the ROIC acts as a charge to voltage signal converter and contains all readout circuitry. The layers are precisely aligned and then joined together at each pixel through indium bump bonds. The fact that the layers are kept separate allows for a unique strength of HCDs --- separate optimization for each layer. The silicon absorption layer is optimized for very high quantum efficiency across the soft X-ray band pass and allows for detection of X-rays from $0.2 - 20$ keV \cite{HCD_QE}. This is thanks to the ability to use high resistivity silicon in the absorption layer and thus achieve depletion regions of greater than 100 microns. Since the first successful demonstration in 2007 \cite{hcd_aluminum}, X-ray HCDs have commonly included a thin layer of aluminum deposited directly on top of the silicon absorbing layer to block optical light, which would otherwise become a large background source. Meanwhile, the separate ROIC can be optimized for fast readout while minimizing read noise. The ROIC can also be modified to increase functionality with the addition of on-chip based signal processing and support for multiple simultaneous readout formats. Figure \ref{fig:hcd} shows a schematic of an X-ray HCD.

\noindent Compared to modern CCDs, HCDs have a number of advantages. These include:

 \textbf{Faster Readout Rate:} The orders-of-magnitude increase in readout speed that HCDs provide over CCDs mitigates the issues of pile-up. Pile-up occurs when multiple X-ray photons strike a single pixel between read-outs, therefore counting as a single event during a read and degrading information about photon energy. As already mentioned, pile-up would be a major problem for CCDs operating on a large effective area mission like \textit{Lynx}. In addition to the much faster full frame readout rates of HCDs, the ability to selectively read out certain pixels further improves effective read speeds.

 \textbf{Radiation Hardness:} Due to HCDs directly reading every pixel in the array, they are much less susceptible to radiation damage than CCDs. Proton displacement damage has been shown to significantly degrade a CCD's charge transfer efficiency (CTE) over time due to the bucket brigade readout scheme \cite{CTE}. Instead of transferring charge over multiple centimeters of silicon, HCDs need only transfer charge through the absorber thickness ($\sim$ several hundred microns), therefore granting them excellent radiation hardness ($\sim 50$ krads)\cite{rad_hard}. 

 \textbf{Resistance to Micrometeoroid Damage:} In addition to radiation damage, CCDs also suffer from micrometeoroid damage (especially front illuminated CCDs with exposed gate structures). CCDs on current X-ray missions have been shown to undergo serious damage to these gate structures from micrometeoroids \cite{MM_damage}. The damage has the potential to impact both the CCD gate structures directly or affect read out of columns through damaged pixels. HCDs are expected to be much more robust to micrometeoroid damage. This is because they do not have any exposed gates, in addition to each pixel containing separate read out architecture that will prevent damage from one pixel affecting whole columns. 

 \textbf{Lower Power:} The readout technique of CCDs requires large capacitive loads in the parallel clock gates and therefore results in large power draws. Due to their lower capacitance CMOS switches, order of magnitude reduction in power requirements can be achieved with HCDs \cite{low_power}. For example, \textit{Swift} XRT uses 8.4 Watts to perform a full readout of the CCD, whereas a $1024\times1024$ H1RG HCD uses only $\sim 200$ mW for all bias generation and readout\cite{hcd_2}. The low power requirements of HCDs enable mission designs to use large arrays of small pixels operating at high frame rates.
 
\subsection{HCD measurements with cryogenic SIDECAR$^{\text{TM}}$}
\label{sec:cryo}

The Teledyne SIDECAR$^{\text{TM}}$ application-specific integrated circuit (ASIC) is used to provide clock and bias signals to HxRG (here x refers to the size of the ROIC array in multiples of 1024 pixels) detectors while also performing chip programming, signal amplification, analog to digital conversion, and data buffering. We recently acquired a cryogenic SIDECAR$^{\text{TM}}$ ASIC from TIS for use with our HxRG detectors. When cooled to $<$ $\sim 200$ K, the cryogenic SIDECAR$^{\text{TM}}$ promises improved noise performance compared to room temperature operation due to the reduction of thermal voltage fluctuations in the ASIC during integration. PSU has reported on previous testing cycles of HCDs with a room temperature SIDECAR$^{\text{TM}}$ ASIC \cite{hcd_1}; in this proceeding we report initial measurements of an H2RG X-ray HCD with a cooled cryogenic SIDECAR$^{\text{TM}}$.

The H2RG detector used for these measurements is a modified engineering-grade device fabricated by TIS. This detector has a $2048\times2048$ pixel ROIC with 18 $\mu$m pitch and a $1024\times1024$ absorber layer with 36 $\mu$m pitch pixels. Every absorber pixel has only one ROIC pixel bonded to it, meaning the effective pitch of the hybridized detector is 36 $\mu$m. This nonstandard layout was designed to reduce the effects of interpixel capacitance (IPC), whereby a parasitic capacitance exists between neighboring pixels and leads to signal sharing between nearby pixels. Previous generation H1RG detectors tested in our lab had substantial IPC, with up to $ \sim 10$\% of the signal shared with a given neighborhood pixel. The modified H2RG was shown to be a significant improvement and limited IPC to $\sim 1.8$\% (note: implementation of CTIA amplifiers in the ROICs of even more recent devices has resulted in no measurable IPC, as discussed later). Operated with the room temperature SIDECAR$^{\text{TM}}$, this H2RG was measured to have a read noise of 16.31 e$^-$ and had modest energy resolution of 369 eV (6.3\%) at 5.9 keV and 156 eV (10.5\%) at 1.5 keV. The best energy resolution measured with the room temperature ASIC for any HxRG detector is 248 eV (4.2\%) at 5.9 keV. \cite{hcd_char}

The new measurements of the H2RG were made using the PSU cube test stand, which is a light-tight vacuum chamber. The chamber is evacuated to a pressure of $10^{-6}$ torr before cooling the copper cold finger mounted detector to 130 K. A copper cold strap connecting the cold finger to the SIDECAR$^{\text{TM}}$ cools the ASIC to $\sim 185$ K. Figure \ref{fig:cube} shows a picture of the interior of the cube test chamber with the H2RG and cryogenic SIDECAR$^{\text{TM}}$ visible. Further signal processing and amplification are performed with a TIS supplied SIDECAR acquisition module (SAM) card also mounted in the chamber. Readout scheme (including reset method and integration time) is controlled through TIS supplied HxRG software. Our data collection method is to perform three detector resets followed by a ``ramp'' of image taking that typically includes 100 images before another set of resets. Each image is non-destructively read out with either 32 or 4 parallel output channels every 1.48 s or 10.65 s respectively. The raw ramped images are processed using a software CDS algorithm, and then with horizontal and channel offset corrections. The latter involves fitting a separate Gaussian to all pixels in each individual output channel and subtracting from each channel its centroid value; the same process is then repeated with all pixels in each row.

A $^{55}$Fe source was used to produce 5.9 keV and 6.5 keV X-rays from Mn K$\alpha$ and K$\beta$ emission, while two $^{210}$Po alpha particle sources were utilized to fluoresce several lower energy X-ray lines: oxygen, magnesium, and aluminum at 0.53 keV, 1.25 keV, and 1.49 keV respectively. X-ray data collected with the H2RG were used to create X-ray energy spectra by applying thresholding and grading to each X-ray event. Each event was required to be above a $\sim 5\sigma$ primary threshold (where $\sigma$ is the RMS read noise). Pixels above this threshold in Digital Number (DN) have their immediately neighboring regions saved in an event list. A $\sim 3\sigma$ secondary threshold is applied to the $3 \times 3$ event neighborhood and any pixel above the threshold is added to the event. Next, \textit{Swift}/XRT grade definitions \cite{swift_XRT} are applied by finding the number and position of pixels that exceed the secondary threshold. In this experiment we used only single pixel events (Grade 0) and thus only those events that had no neighboring pixels above the secondary threshold were selected for inclusion in the spectra.

\begin{figure}[h]
   \begin{center}
    \begin{minipage}[b]{0.45\textwidth}
    \includegraphics[width=\textwidth, clip, trim=1.5cm 12cm 2cm 3.5cm]{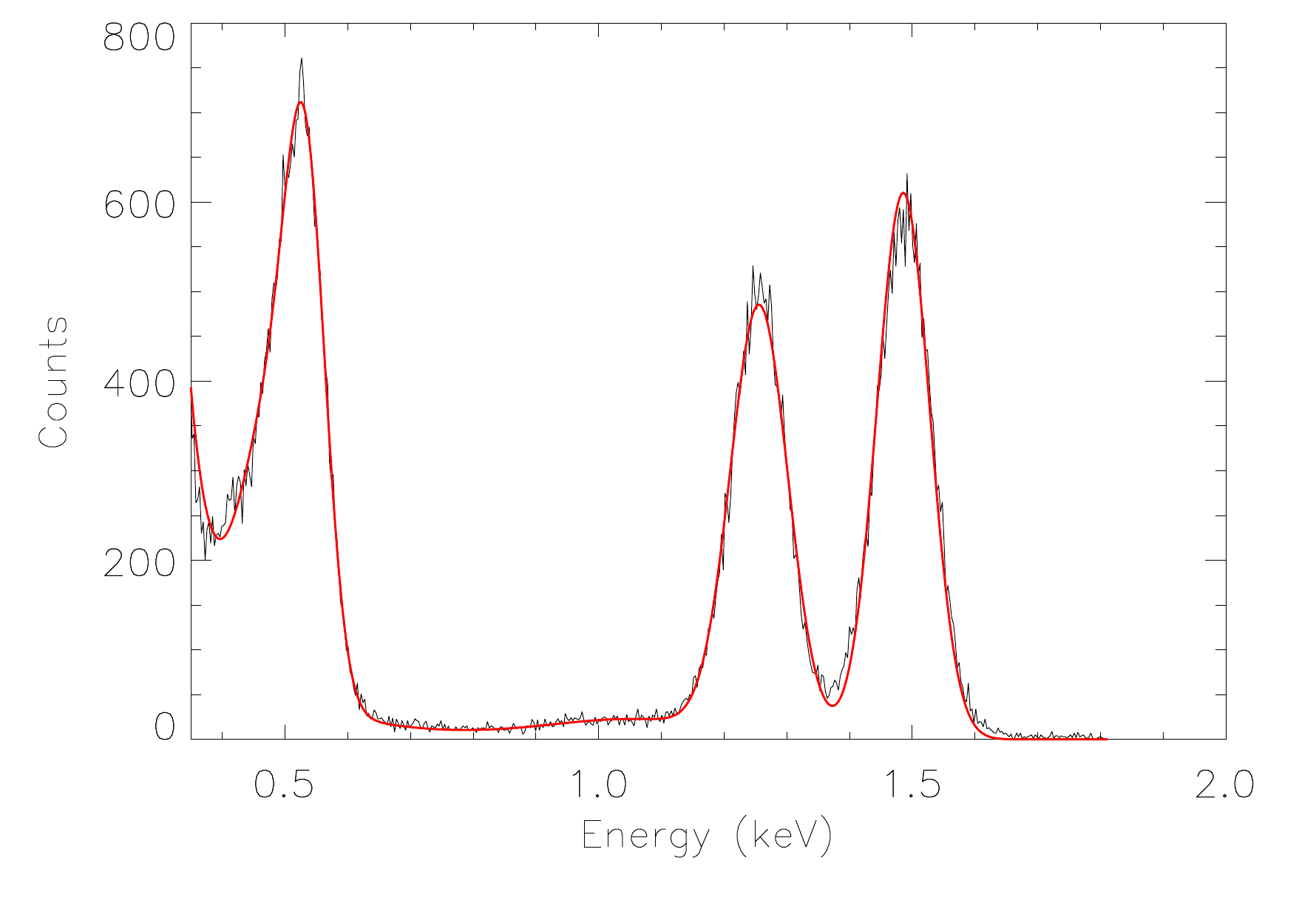}
    \caption{\label{fig:h2rg_lowE} A H2RG spectrum with O (0.53 keV), Mg (1.24 keV), and Al (1.49 keV) lines. A fit to the data is shown in red. }
  \end{minipage}
  \hfill
  \begin{minipage}[b]{0.45\textwidth}
    \includegraphics[width=\textwidth, clip, trim=1.5cm 12cm 2cm 3.5cm]{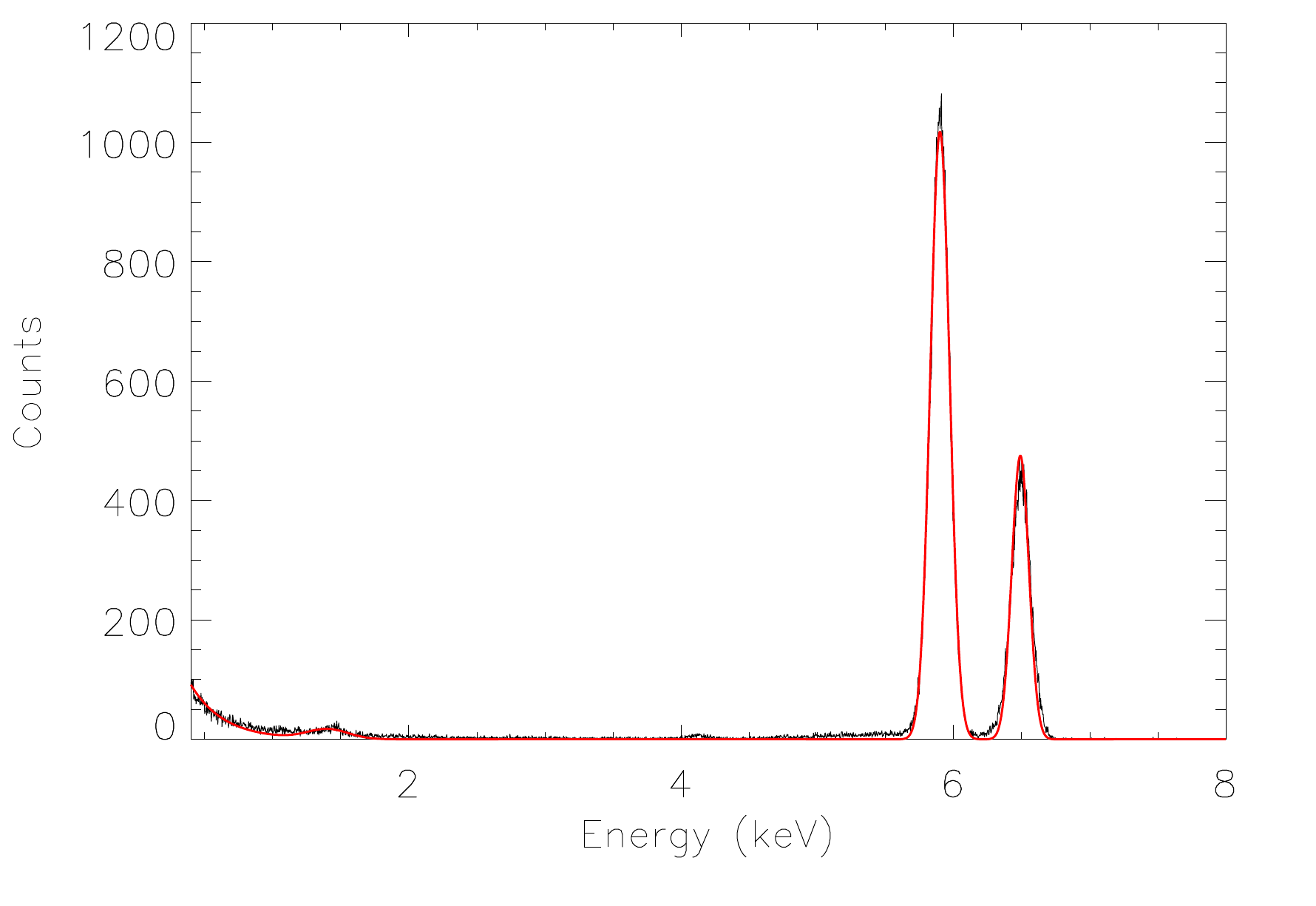}
    \caption{\label{fig:h2rg_fe55} A H2RG spectrum showing the Mn K$\alpha$ and K$\beta$ lines at 5.9 and 6.49 keV. A fit to the data is shown in red. }
  \end{minipage}
   \end{center}
   \end{figure}

Single pixel H2RG spectra showing the Mn, Al, Mg, and O lines are shown in Figures \ref{fig:h2rg_lowE} and \ref{fig:h2rg_fe55}. The $^{55}$Fe data were taken in 32 channel mode, while the lower energy data were taken in 4 channel mode. The measured (FWHM) energy resolution for each of these lines is reported in Table \ref{tab:h2rg_deltaE}. Notably, we achieve 2.7\% and 6.8\% energy resolution at 5.9 keV and 1.5 keV respectively --- both of these are the best values ever reported using an X-ray HCD. Further, the O line at 0.53 keV is well separated from the noise peak and still maintains modest $93\pm4$ eV (17.7\%) FWHM energy resolution.

Read noise was also measured for the H2RG with the cryogenic SIDECAR$^{\text{TM}}$. This was obtained by taking a ramp of dark images at cryogenic temperatures (130 K) and finding the standard deviation of each dark frame histogram. The read noise is then calculated as the mean value of this standard deviation across all dark images. The resulting read noise using the H2RG in 32 channel mode was found to be $6.78$ e$^- \pm$ $0.07$ e$^-$. This is the lowest read noise obtained for any X-ray HxRG detector.

\begin{table}[ht]
\begin{center}   
\def\arraystretch{1.5}    
\begin{tabular}{l|l|l|l|l} 
\hline
\rule[-1ex]{0pt}{3.5ex}  \textbf{Energy (keV)} & 0.53 (O K$\alpha$) & 1.25 (Mg K$\alpha$) & 1.49 (Al K$\alpha$) & 5.90 (Mn K$\alpha$) \\
\hline
\rule[-1ex]{0pt}{3.5ex}  \boldmath$\Delta E$ \textbf{(eV)} & 93 $\pm$ 4 & 111 $\pm$ 3 & 102 $\pm$ 3    & 157 $\pm$ 4  \\

\end{tabular}
\medskip
\caption{H2RG energy resolution with cryogenic SIDECAR$^{\text{TM}}$ (Grade 0 events). Data for Mn K$\alpha$ taken in 32 channel mode while O, Mg, and Al data taken in 4 channel mode. Values are FWHM.} 
\label{tab:h2rg_deltaE}
\end{center}
\end{table}

\section{Small Pixel HCDs}
\label{sec:sp}

\begin{figure}
\centering
\vspace{0cm}
\hspace*{0 cm}\includegraphics[clip, width=.50\linewidth]{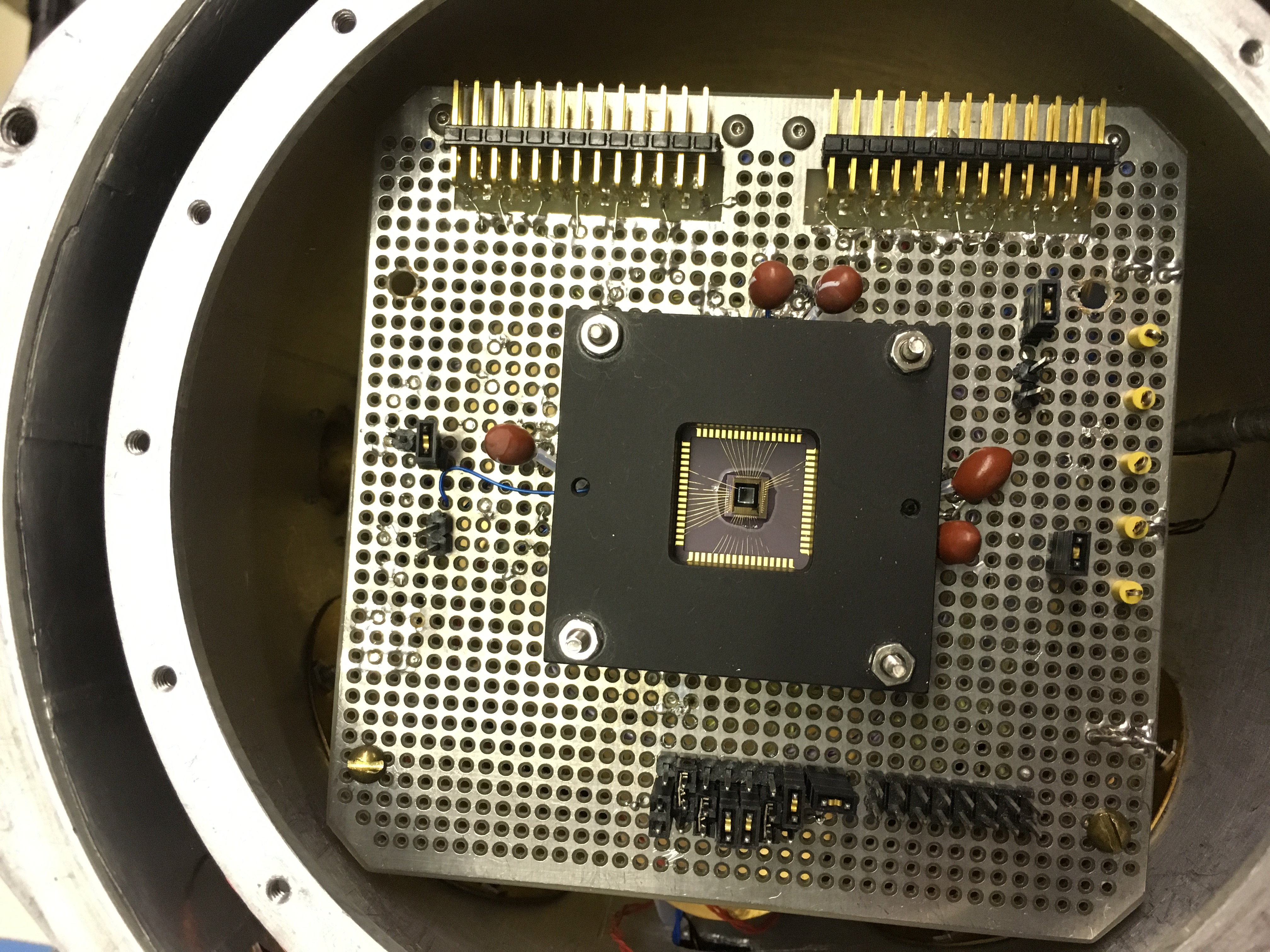}
\vspace{.2cm}
\caption{A small pixel detector (18567) seated in breadboard inside testing dewar. The breadboard measures 10 cm x 10 cm.} 
\label{fig:sp_dewar}
\end{figure}

The small pixel hybrid CMOS detectors are new prototype Si X-ray HCDs that were designed as part of a collaboration between PSU and TIS and fabricated by TIS. The goal of the prototype design was to satisfy the small pixel size and high frame rate needs of large effective area missions concepts like \textit{Lynx}, while also improving detector read noise to $< 4$ e$^-$. The $128\times128$ pixel prototypes have a $12.5$ micron pixel pitch and a depletion depth of $100$ microns. While the small area of the detectors limits practical application in its current form, the design can easily be scaled up to accommodate abuttable 4k $\times$ 4k formats (or smaller sizes). These engineering-grade detectors currently have a fixed readout speed of approximately 10 Hz, though firmware updates to the readout board allow for speeds up to 1000 Hz. Figure \ref{fig:sp_dewar} shows one of the small pixel HCDs inside a testing dewar. TIS has fabricated four mostly identical small pixel prototype HCD detectors, the properties of which are summarized in Table \ref{table:sp_properties}.

\begin{table}[h]
\def\arraystretch{1.1}
\setlength\arrayrulewidth{1pt}  
\centering
\rowcolors{3}{lightgray}{}
\begin{tabular}{  c c c c c }
\multicolumn{5}{ c }{\textbf{Small Pixel Detector Properties} } \\
\toprule
Detector \# & Pixel Pitch ($\mu$m) & Dimensions (pixels) & Depletion Depth ($\mu$m) & Al Filter Thickness ( $\angstrom$ ) \\ \midrule
18566 & 12.5 & $128\times128$ & 100 & 0 \\
18567 & 12.5 & $128\times128$ & 100 & 500 \\
18568 & 12.5 & $128\times128$ & 100 & 500 \\
18569 & 12.5 & $128\times128$ & 100 & 0 \\
\end{tabular}
\caption{Properties of the small pixel HCDs.}
\label{table:sp_properties}
\end{table}

\noindent Below we discuss the key features of the small pixel HCDs:

\textbf{CTIA Amplifier}: The small pixel detectors use a Capacitive Transimpedance Amplifier (CTIA) in each pixel. This input amplifier is key to eliminating interpixel coupling. The HAWAII HCDs use a source follower amplifier; when the input gate voltage of a pixel changes it is capacitively coupled to neighboring pixels and therefore results in IPC that degrades detector performance. The CTIA instead holds input gate voltages constant during integration and thus effectively eliminates the problem of IPC. The CTIA was successfully implemented in the Speedster-EXD and was shown to result in no measureable IPC \cite{speedster1}.

\textbf{In-Pixel CDS Subtraction}: Following the CTIA in the ROIC, the small pixel detector design can include an in-pixel CDS subtraction circuit. This circuit enables the subtraction of the variable baseline voltage level associated with a reset and therefore is used to cancel the reset kTC noise. The subtraction is also useful for reducing 1/f noise and is carried out on-chip before further amplification. 

\begin{figure}[t]
   \begin{center}
    \begin{minipage}[b]{0.45\textwidth}
    \includegraphics[width=\textwidth]{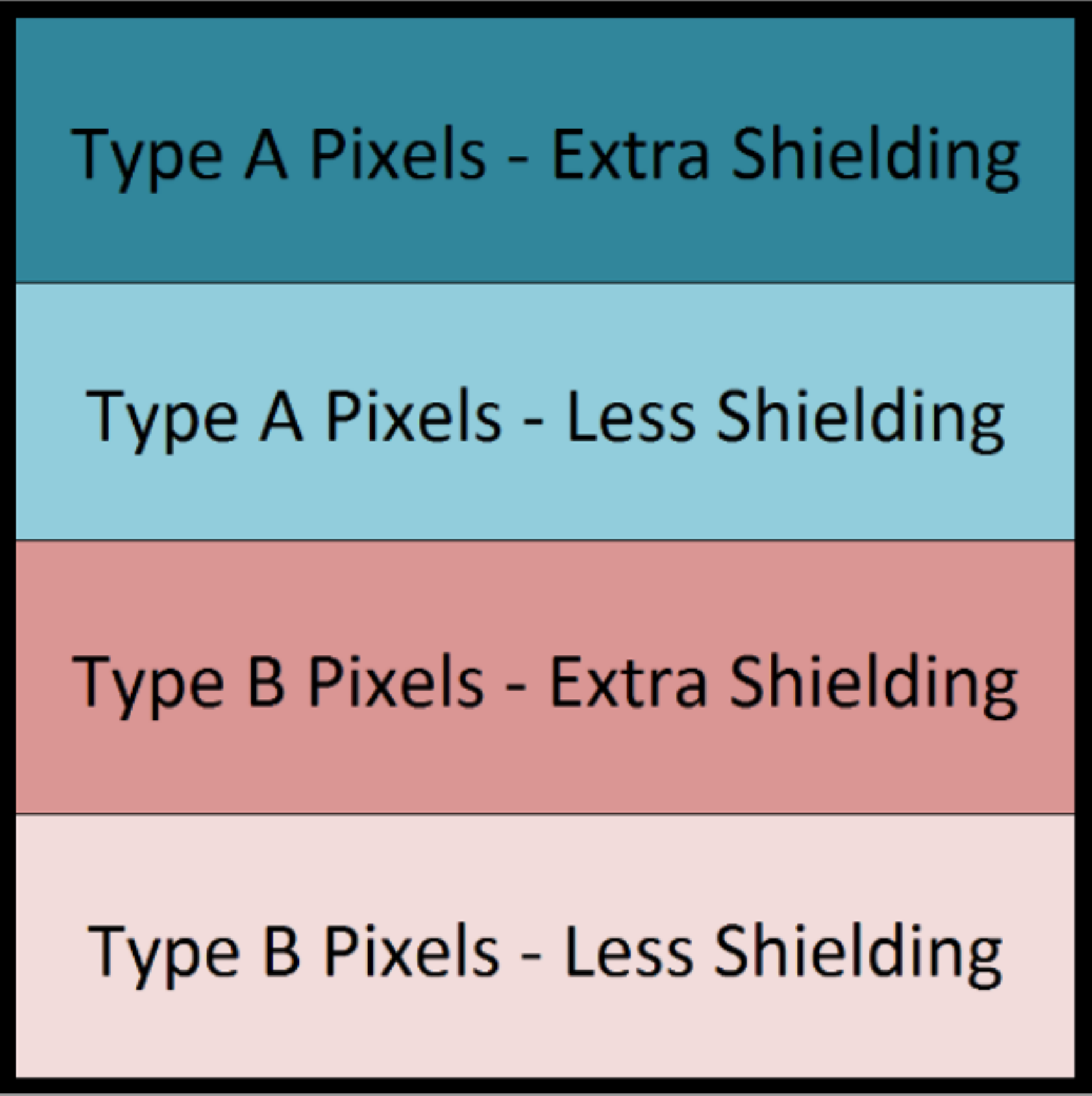}
    \smallskip
    \caption{The small pixel test chip array. Type A pixels do not contain in-pixel CDS circuitry, while type B pixels do. Each band is 32 rows $\times$ 128 columns.} 
\label{fig:sp_bands}
  \end{minipage}
  \hfill
  \begin{minipage}[b]{0.5\textwidth}
    \includegraphics[width=\textwidth]{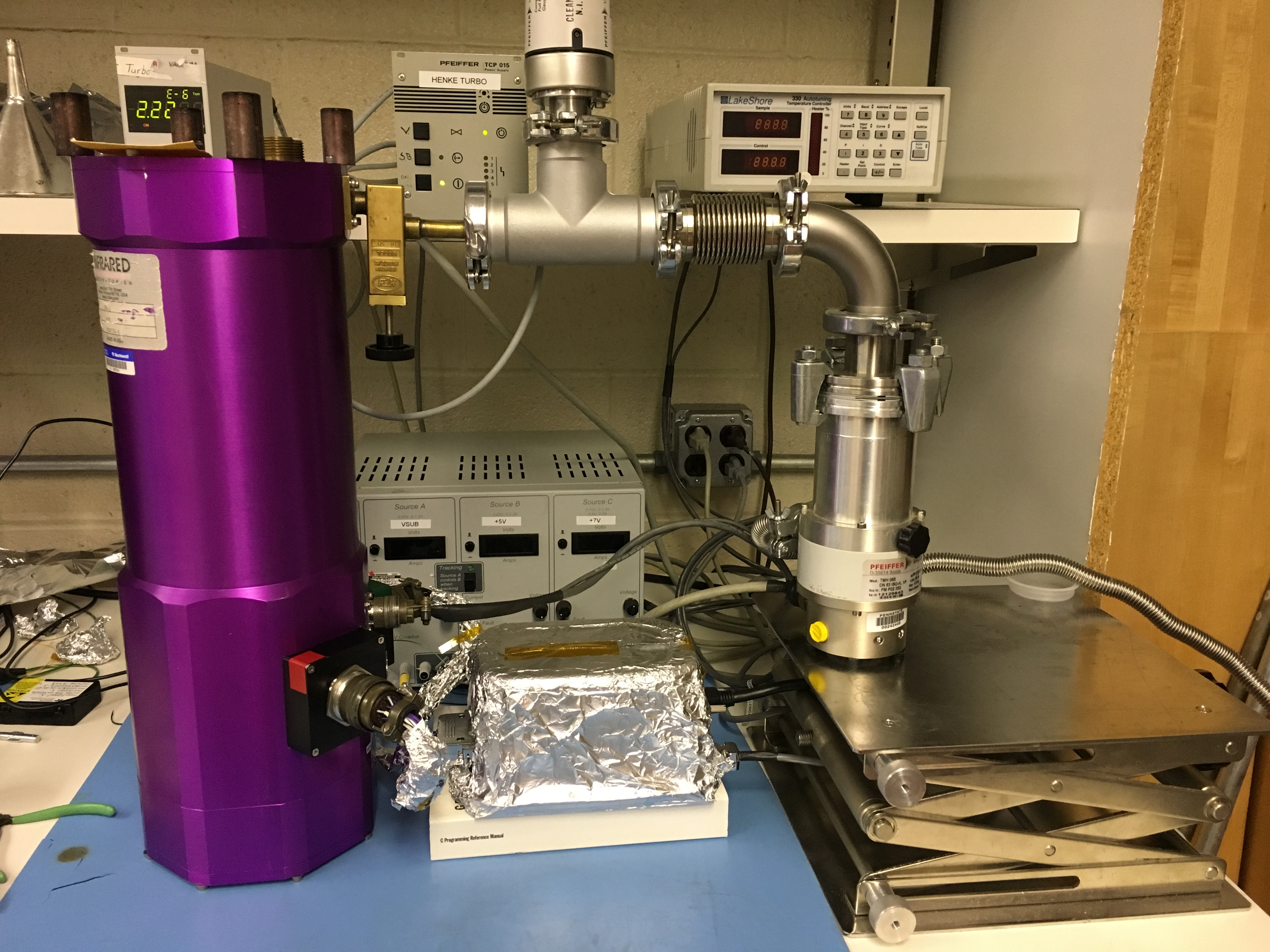}
    \smallskip
    \caption{The small pixel test setup. The HDL-5 dewar is on the left and the vacuum system components are on the right. In the middle is the detector interface block (DIB) board and power supply.} 
\label{fig:teststand}
  \end{minipage}
   \end{center}
   \end{figure}

\textbf{Pixel Layout:} The in-pixel CDS circuit is not implemented for every pixel of the small pixel HCD devices. In fact the prototype devices were fabricated with ``banded arrays'' which allow for the testing of multiple different types of pixels. Each of the four small pixel HCDs is divided into four different bands. The top two bands, or the first 64 rows, contain ``Type A'' pixels and do not include the in-pixel CDS circuitry. The bottom two bands, or the last 64 rows, contain ``Type B'' pixels with on-chip CDS capability. The first band of each pixel type contains a pixel design that includes additional shielding in the ROIC designed to further reduce pixel crosstalk. The lower 32 rows of each pixel type contain normal pixels that do not include this extra shielding. This layout is shown in Figure \ref{fig:sp_bands}, and allows for pixel performance to be compared and evaluated. 

\subsection{Small pixel experimental setup}

A custom IR Labs HDL-5 dewar was used to cool the small pixel detectors and expose them to X-rays. The detector is mounted in a socket on the dewar breadboard as shown in Figure \ref{fig:sp_dewar}, which is then secured inside the vacuum chamber portion of the HDL-5. The detector socket has been designed with a hole in the bottom to allow for the cold finger to make contact with and cool the detector. The dewar breadboard mostly serves to facilitate routing of all external dynamic and analog signals to connections on the dewar exterior. However due to I/O limitations, seven static digital signals are set manually by means of jumpers on the breadboard. The breadboard also includes a low-pass filter on the substrate voltage connection as well as filtering capacitors for the supply pins. Two separate dewar connectors for analog and digital lines carry all signals to an interface board (Detector Interface Block, DIB) developed by TIS. The detector is controlled using the DIB board, a Matrox frame grabber, and modified TIS Speedster-EXD software that allows for control over which type of pixel is read out. The HDL-5 dewar in the full test setup is shown in Figure \ref{fig:teststand}. 

The dewar was evacuated to a pressure of $\sim 5 \times 10^{-6}$ torr, which allows for safe cooling of the detectors to 150 K and thus minimization of dark current noise contributions. Temperature was precisely controlled with a Lake Shore temperature controller in concert with a platinum RTD and heater mounted near the dewar coldfinger. $^{55}$Fe was used as an X-ray source, producing the familiar Mn K$\alpha$ and Mn K$\beta$ X-rays at $5.90$ keV and $6.49$ keV respectively. The detectors are highly depleted and run with 15 V applied to the substrate.

\subsection{Small pixel analysis and results}

The HDL-5 dewar was used to obtain read noise and energy resolution measurements for all four small pixel detectors for both pixel types and shielding configurations. The exception is detector \#18566, for which only type B data were acquired before a detector package wire bond was damaged. Data from A and B type pixels must be acquired separately due to the different output modes for each pixel type. Type A pixels output a reset and signal frame for each exposure, whereas type B pixels only output a single CDS subtracted frame. For type A pixels, the signal frame was subtracted from the reset frame to create a pseudo-CDS subtracted image. For type B pixels, a bias frame was obtained from the average of 1000 dark frames with no X-rays, which was then subtracted from each image to create a bias subtracted frame. For both pixel types the resulting subtracted frames were then run through a boxcar smoothing algorithm.

Detector gain was calculated using 10,000 exposures with the $^{55}$Fe source, which allows for creation of X-ray spectra with Mn lines of known energy. After image subtraction, primary and secondary thresholds were applied to locate and apply initial grading to events. Because of the small pixel size and large depletion depths of these detectors, very few single pixel events are recorded and the vast majority of events spread to 4-6 pixels. Event candidates are graded using the same scheme as described in Section \ref{sec:cryo}, except that \textit{Chandra} ACIS grades\cite{ACIS} were used instead of \textit{Swift}/XRT grades due to the increased charge spreading. All plausible X-ray event configurations were included for this experiment due to the lack of single or doubly split pixel events.

\begin{figure}[h]
\begin{subfigure}{0.48\textwidth}
\centering
\includegraphics[clip, trim=2.5cm 1.8cm 3.2cm 2.9cm, height=0.20\textheight]{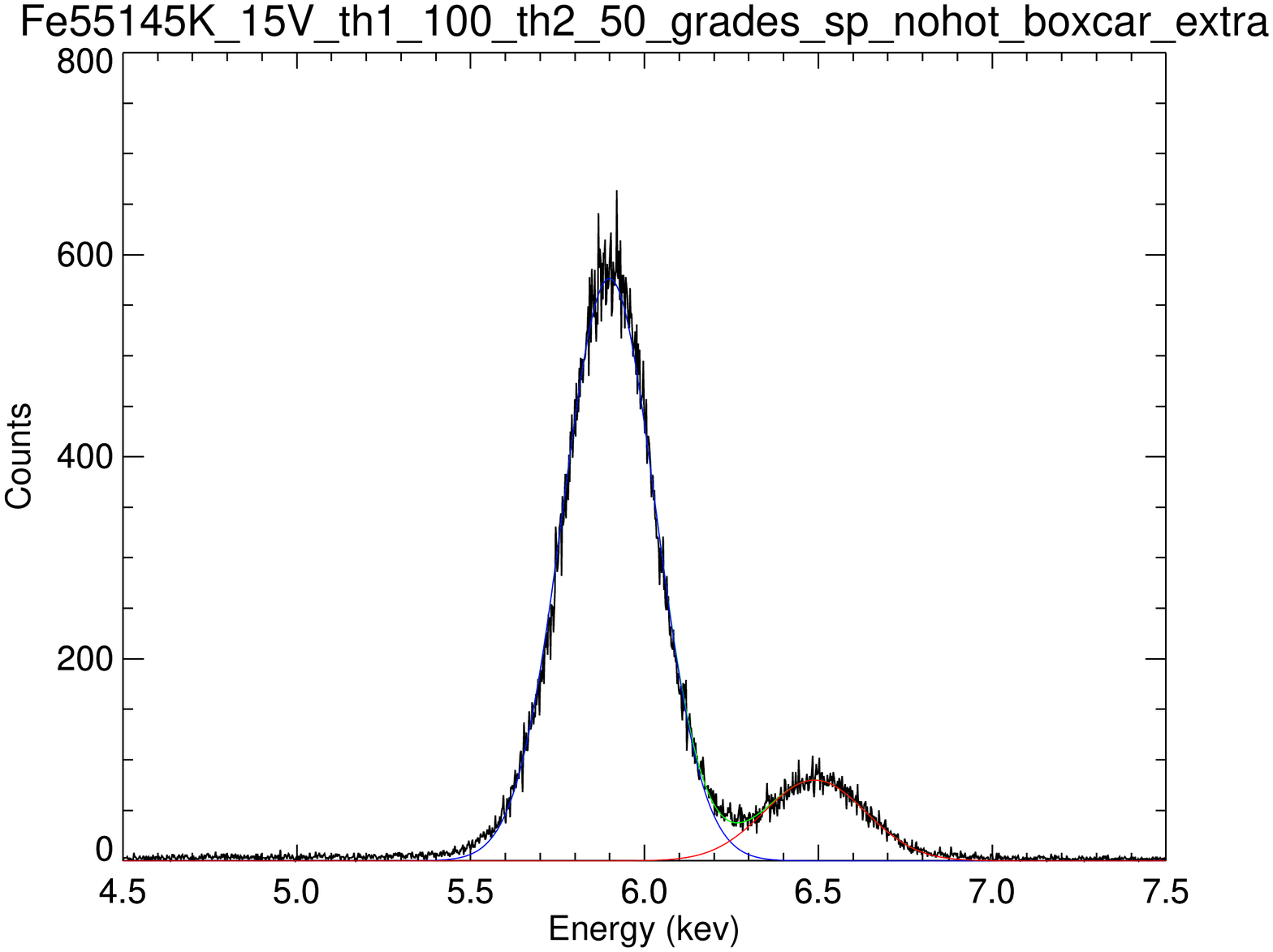}
\caption{Type A (no in-pixel CDS) - extra shielding} \label{fig:a}
\end{subfigure}\hspace*{\fill}
\hspace{\fill}
\begin{subfigure}{0.48\textwidth}
\centering
\includegraphics[clip, trim=2.5cm 1.8cm 3.2cm 2.9cm, height=0.20\textheight]{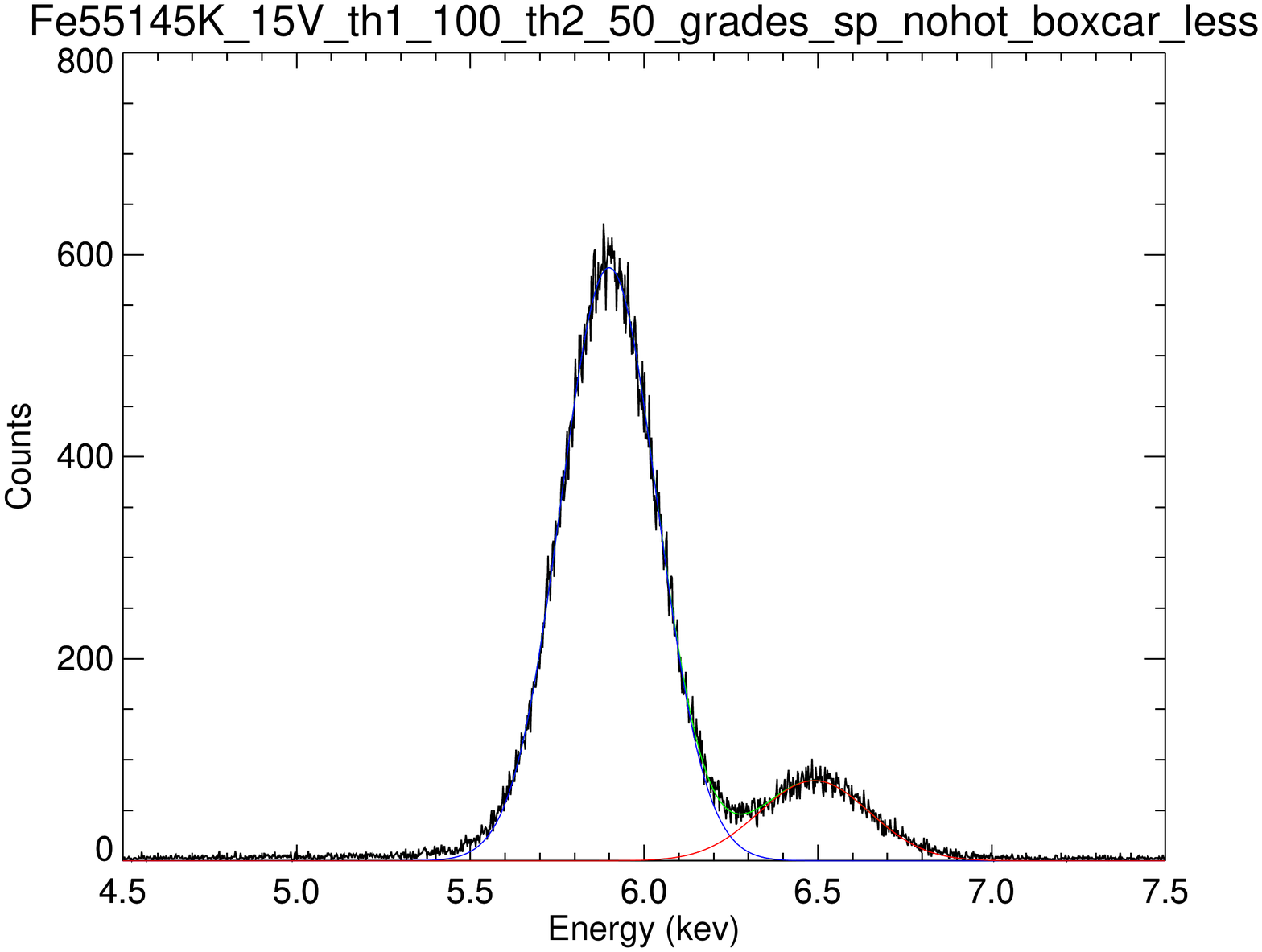}
\caption{Type A (no in-pixel CDS) - less shielding} \label{fig:b}
\end{subfigure}
\medskip
\begin{subfigure}{0.48\textwidth}
\centering
\includegraphics[clip, trim=2cm 1.8cm 3.2cm 2.9cm, height=0.20\textheight]{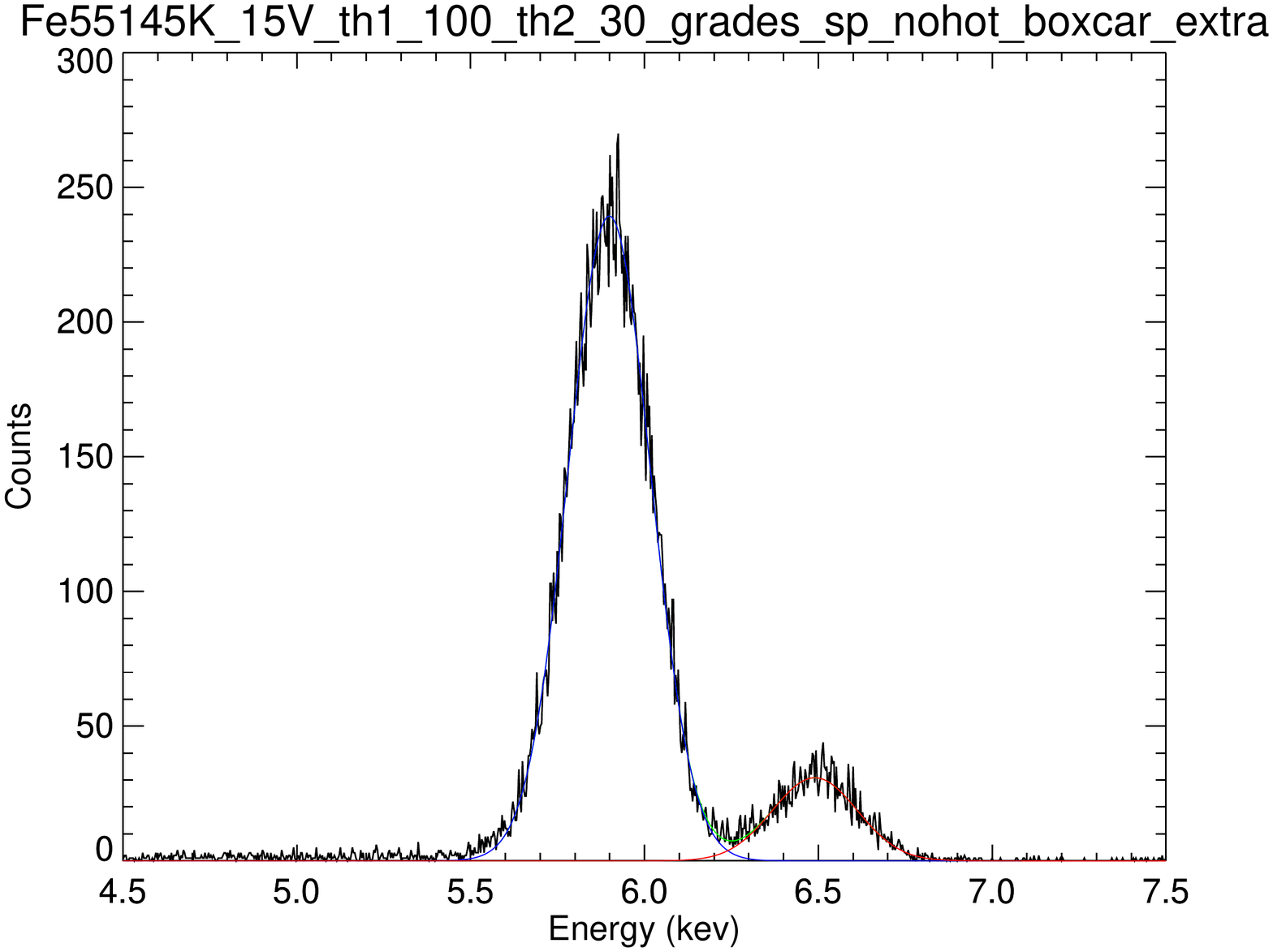}
\caption{Type B (in-pixel CDS) - extra shielding} \label{fig:c}
\end{subfigure}\hspace*{\fill}
\begin{subfigure}{0.48\textwidth}
\centering
\includegraphics[clip, trim=2cm 1.8cm 3.2cm 2.9cm, height=0.20\textheight]{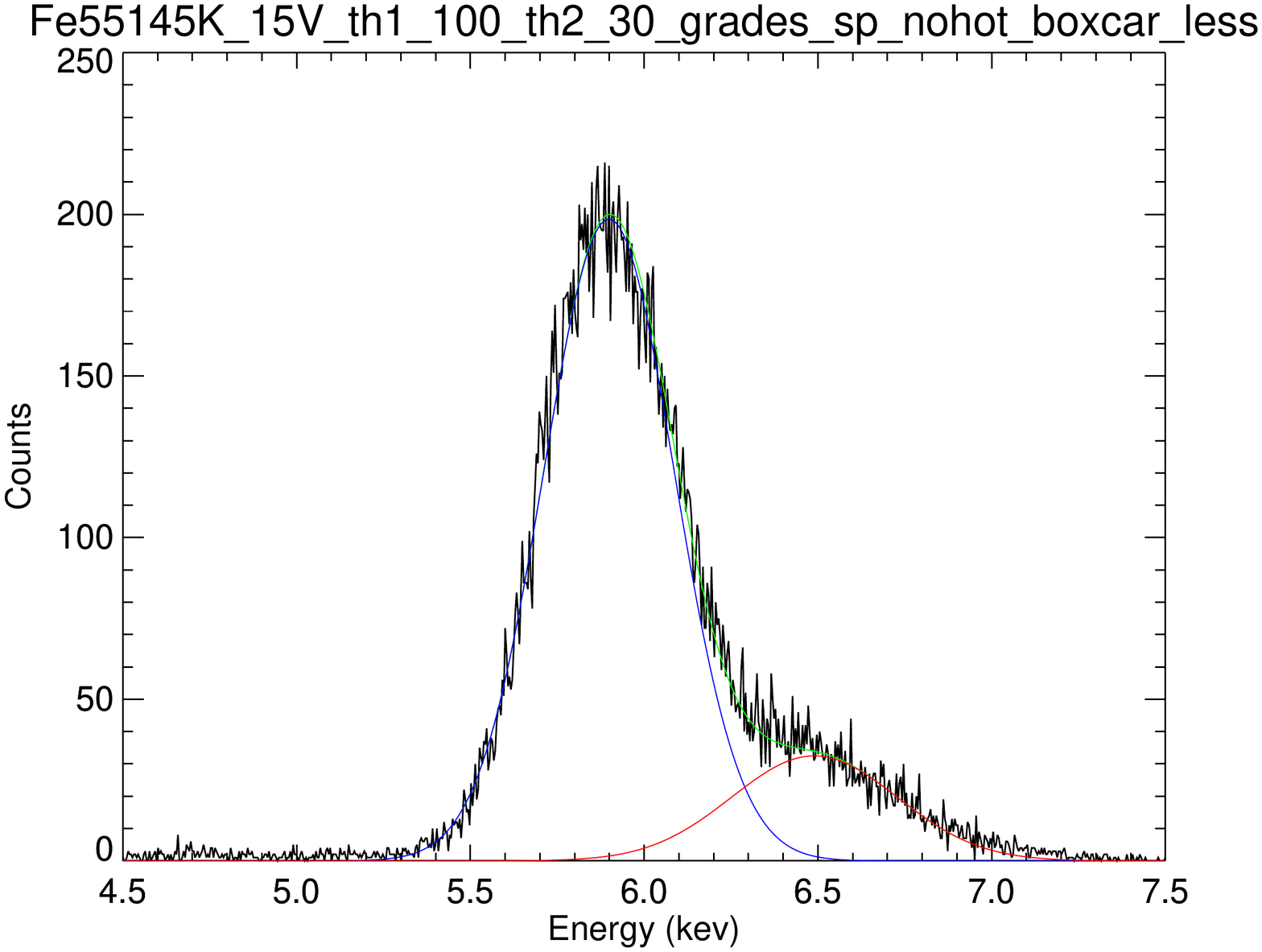}
\caption{Type B (in-pixel CDS) - less shielding} \label{fig:d}
\end{subfigure}

\caption{$^{55}$Fe X-ray spectra for detector \#18568, showing the Mn K$\alpha$ and Mn K$\beta$ lines at $5.90$ keV and $6.49$ keV. The K$\alpha$ fit is shown in blue, the K$\beta$ in red, and the total fit in green.} \label{18568_spectra}
\end{figure}

The system gain was evaluated using the resulting Mn X-ray events. Graded spectra were used for each pixel type and shielding level to obtain a gain conversion at 150 K. Gaussian fits were used to fit the Mn K$\alpha$ and K$\beta$ peaks for each spectrum and obtain centroid positions. The e$^-$/DN conversion was  obtained using the standard conversion factor for silicon of $3.65$ eV/e$^-$ \cite{janesick}. The resulting gain conversion factors are summarized in Table \ref{table:gain}. The Gaussian fit also allows for a determination of the energy resolution at 5.9 keV through measurement of the full width at half maximum (FWHM) of the K$\alpha$ peak. The energy resolution at 5.9 keV for each detector and pixel type is summarized in Table \ref{table:deltaE}, and example spectra for one small pixel detector are shown in Figure \ref{18568_spectra}. Type B (in-pixel CDS) pixels with extra shielding consistently have the best energy resolution, with the best measured value being 278 eV (4.7\%) at 5.9 keV. It should be noted that these measurements do not take into account pixel-to-pixel gain variation, which is thought to be a major contributing factor to the energy resolution in these detectors. 

\begin{table}[t]
\setlength{\arrayrulewidth}{0.75pt}
\rowcolors{4}{white}{lightgray}
\centering
\def\arraystretch{1.5}
\begin{tabular}{ l| l| l| l| l }
\multicolumn{5}{ c }{\textbf{Small Pixel Detector Gain} } \\
\toprule
\multicolumn{1}{c}{} & \multicolumn{2}{ c }{Type A} &  \multicolumn{2}{ c }{Type B} \\
\multicolumn{1}{c}{Detector} & \multicolumn{1}{c}{Extra} & \multicolumn{1}{c}{Less} & \multicolumn{1}{c}{Extra} & \multicolumn{1}{c}{Less} \\ \midrule
 18566 & --- & --- & $0.97$ e$^-$/DN&  $1.01$ e$^-$/DN \\ \midrule
 18567 & $0.57$ e$^-$/DN & $0.60$ e$^-$/DN& $0.94$ e$^-$/DN& $1.01$ e$^-$/DN\\ \midrule
 18568 & $0.58$ e$^-$/DN & $0.59$ e$^-$/DN & $0.95$ e$^-$/DN & $1.01$ e$^-$/DN \\ \midrule
 18569 & $0.58$ e$^-$/DN &  $0.59$ e$^-$/DN & $0.96$ e$^-$/DN & $1.01$ e$^-$/DN \\
\end{tabular}
\caption{Measured gain for small pixel detectors with each pixel type and shielding level.}
\label{table:gain}
\end{table}

\begin{table}[b]
\setlength{\arrayrulewidth}{0.75pt}
\rowcolors{4}{}{lightgray}
\centering
\def\arraystretch{1.5}
\begin{tabular}{ l|l|l|l|l }
\multicolumn{5}{ c }{\textbf{Small Pixel Energy Resolution at 5.9 keV} } \\
\toprule
\multicolumn{1}{c}{} & \multicolumn{2}{ c }{Type A} &  \multicolumn{2}{ c }{Type B} \\
\multicolumn{1}{c}{Detector} & \multicolumn{1}{c}{Extra (eV)} & \multicolumn{1}{c}{Less (eV)} & \multicolumn{1}{c}{Extra (eV)} & \multicolumn{1}{c}{Less (eV)} \\
 \midrule
 18566 & --- & --- & $292 \pm 2$ &  $398\pm 4$ \\
 \midrule
 18567 & $345\pm 5$ & $343 \pm 4$ & $313 \pm 4$ & $426 \pm 8$ \\
 \midrule
 18568 & $314 \pm 2$ & $ 322 \pm 3$ & $278 \pm 4$ & $ 398 \pm 5$ \\
 \midrule
 18569 & $305 \pm 3$ &  $323 \pm 4$ & $286 \pm 3$ & $372 \pm 7$ \\
\end{tabular}
\caption{Measured energy resolution at 5.9 keV for each pixel type and shielding level.}
\label{table:deltaE}
\end{table}

In order to measure the read noise $\sim$ 1000 dark exposures were taken for each pixel type. The resulting cumulative dark frame histograms were then fit with a simple Gaussian to determine the standard deviation, $\sigma$, of the noise distribution. This method for measuring the read noise, in contrast to the one described in Section \ref{sec:cryo}, is insensitive to the number of hot pixels on the detector. Hot pixels are pixels that, due to lattice defects, may contain charge traps with leakage current that therefore appear as persisting or flickering high DN value pixels. Due to these detectors being prototype devices they may contain a higher than average number of hot pixels. Further, pixels near the boundary of larger (e.g. $1024\times1024$) HCDs are known to contain a greater number of defects and thus a large percentage of the pixels on these $128\times128$ detectors are ``at risk'' edge pixels. Using the fitting method, the resulting read noise measurements are summarized in Table \ref{table:readnoise}. The best measured read noise is 5.54 e$^-$, found in a type A pixel with less shielding.

\begin{table}[h]
\setlength{\arrayrulewidth}{0.85pt}
\rowcolors{4}{}{lightgray}
\centering      
\def\arraystretch{1.5}
\begin{tabular}{ l|l|l|l|l }
\multicolumn{5}{ c }{\textbf{Small Pixel Detector Noise} } \\
\toprule
\multicolumn{1}{c}{} & \multicolumn{2}{ c }{Type A} &  \multicolumn{2}{ c }{Type B} \\
\multicolumn{1}{c}{Detector} & \multicolumn{1}{c}{Extra (e$^-$)} & \multicolumn{1}{c}{Less (e$^-$)} & \multicolumn{1}{c}{Extra (e$^-$)} & \multicolumn{1}{c}{Less (e$^-$)} \\
 \midrule
 18566 & --- & --- & $6.51 \pm 0.05$ &  $7.21$ $\pm 0.07$ \\
 \midrule
 18567 & $6.49\pm 0.11$ & $5.97$ $\pm 0.08$ & $6.16 \pm 0.06$ & $7.28 \pm 0.07$ \\
 \midrule
 18568 & $5.57 \pm 0.04$ & $5.54 \pm 0.05$ & $5.85 \pm 0.05$ & $6.66 \pm 0.09$ \\
 \midrule
 18569 & $5.68 \pm 0.07$ &  $5.67 \pm 0.04$ & $6.18 \pm 0.06$ & $6.91 \pm 0.06$ \\
\end{tabular}
\caption{Measured RMS read noise for small pixel detectors using a Gaussian fit.}
\label{table:readnoise}
\end{table}

Overall, pixels with in-pixel CDS (type B) and extra shielding have the best performance. These pixels have moderately better energy resolution (by $\sim 0.5$\%) than any of the other pixel types, while having barely higher ($\sim 0.5$ e$^-$) read noise. Future HCD designs that use the small pixel ROIC architecture should iterate on this pixel type. In general, the measured energy resolutions are somewhat high given the read noise values obtained, likely due to the unaccounted-for effect of pixel-to-pixel gain variation. In addition, the large amount of charge spreading will negatively affect the energy resolution; this could be mitigated with higher substrate voltages. Read noise performance is on the other hand very promising. While not quite meeting the stated goal of $<$ $4$ e$^-$, the best measured value of $5.5$ e$^-$ is the lowest X-ray HCD read noise ever measured. This measured value suffers slightly from inclusion of pixel-to-pixel gain variation.

\section{Other HCD Developments}
\label{sec:other}
Several other initiatives are also ongoing in the X-ray detector lab. The continued development of the Speedster-EXD detectors and an overview of plans for the launch of an X-ray HCD carrying sounding rocket will both be described in the following subsections. There are also efforts to measure the sub-pixel spatial resolution of X-ray HCDs with a mesh experiment \cite{mesh}, and to characterize an X-ray lobster eye optic with an HCD \cite{lobster}, both of which utilize the PSU X-ray beamline. Neither of these efforts are described in this conference proceeding, and instead the interested reader is directed to the referenced proceedings.

\subsection{Speedster-EXD}

\begin{figure}[b]
   \begin{center}
    \begin{minipage}[b]{0.45\textwidth}
    \includegraphics[width=\textwidth]{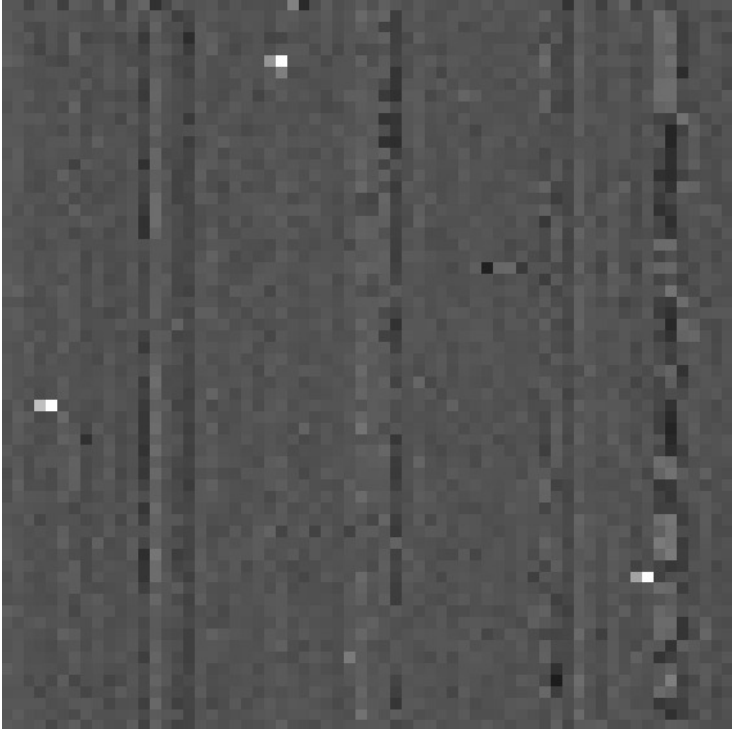}
    \smallskip
    \caption{\label{fig:speedster_fullframe} Full Frame Readout Mode. The comparator is set below the noise floor and thus every pixel in the array is read out. Three multi-pixel $^{55}$Fe events can be seen, in addition to all background pixels.}
  \end{minipage}
  \hfill
  \begin{minipage}[b]{0.45\textwidth}
    \includegraphics[width=\textwidth]{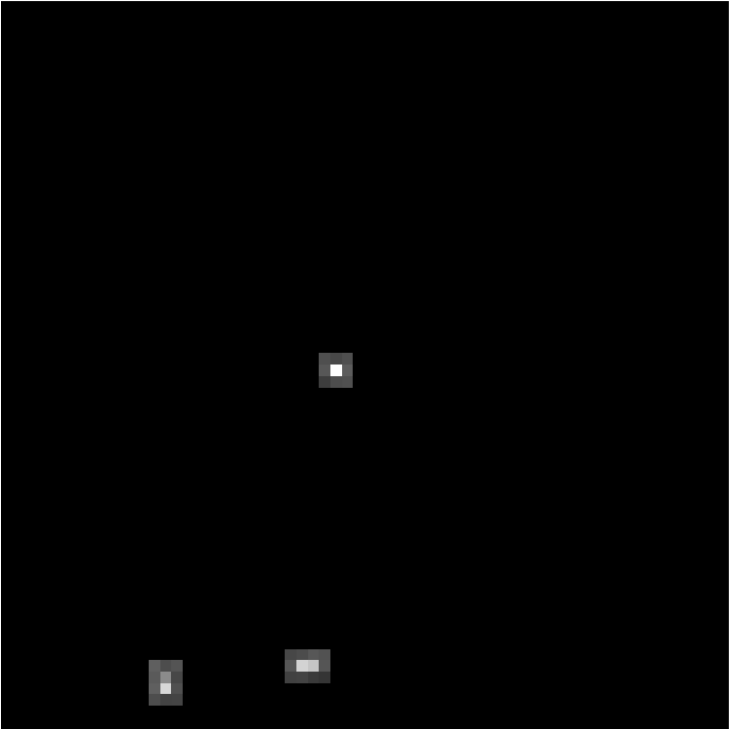}
    \smallskip
    \caption{\label{fig:speedster_sparse} 3x3 Sparse Readout Mode. The comparator is set above the noise floor, allowing read out of only X-ray events and their neighboring pixels. Two multi-pixel and one single pixel $^{55}$Fe X-ray events can be seen, with their 3x3 regions.}
  \end{minipage}
   \end{center}
   \end{figure}

The original Speedster-EXD is a $64\times64$ pixel prototype HCD with 40 micron pixels and additional on-chip circuity that improves upon existing HxRG designs. Like the small pixel detectors, the Speedster-EXD utilizes the IPC eliminating CTIA amplifiers and in-pixel CDS circuity. In addition, the inclusion of a comparator in every pixel opens up a novel readout mode: sparse readout. Sparse readout allows for only those pixels which contain X-ray event charge (above a threshold) to be read out. Typically this is done in 3x3 Sparse Readout Mode, which reads out pixels above the set threshold along with the 3x3 region of pixels surrounding them. A comparison between 3x3 Sparse Readout Mode and Full Frame Readout Mode, in which the comparator is set below the noise floor (thus reading out all $64\times64$ pixels), is shown in Figures \ref{fig:speedster_fullframe} and \ref{fig:speedster_sparse}. The Speedster has the capability to read out at speeds of up to 10 kHz.  

The Speedster-EXD prototype devices have been fully characterized in terms of IPC, read noise, energy resolution, dark current, and pixel-to-pixel gain variation. The best energy resolution was measured to be 206 eV (3.5\%) at 5.9 keV and 172 eV (10.0\%) at 1.49 keV. The pixel-to-pixel gain variation was measured to be as low as $0.80$\% $\pm$ $0.03$\%, and no measureable IPC was detected. For a full characterization of two Speedster-EXD detectors see Griffith et al. (2016) \cite{speedster2}. 

Looking to the future, the next generation Speedster device is currently being developed. This will be a larger format 550 $\times$ 550 pixel device with the same pixel architecture and pixel size. The addition of multiple output lines will allow the device to maintain very high effective frame rates even with many more pixels, while an on-chip analog-to-digital converter will further increase functionality.

\subsection{HCDs on the Water Recovery X-ray Rocket}

The Water Recovery X-ray Rocket (WRXR) is a sounding rocket payload that will launch from Kwajalein Atoll in April 2018. Notably, it will be the first NASA astrophysics sounding rocket payload to attempt water recovery. The payload will be a soft X-ray spectrometer utilizing off-plane reflection gratings\cite{gratings} and a hybrid CMOS detector. The science target of the mission is the Vela supernova remnant, which is a shell type remnant approximately 250 pc distant and with $\sim8^{\circ}$ apparent diameter. The instrument, which has a $3.25^{\circ}$ $\times$ $3.25^{\circ}$ field of view, is optimized to observe 3rd and 4th order OXII and has a resolving power of 40-50 in these lines. WRXR will thus use its moderate spectral resolution and large field of view to probe a large section of the remnant. WRXR will also be an important technology demonstration in a space environment for hybrid CMOS detectors; a successful mission will result in raising the technology readiness level (TRL) of X-ray HCDs to TRL 9.

\begin{figure}[h]
   \begin{center}    
    \includegraphics[clip, trim= 2cm 2cm 2cm 3cm, width=\textwidth]{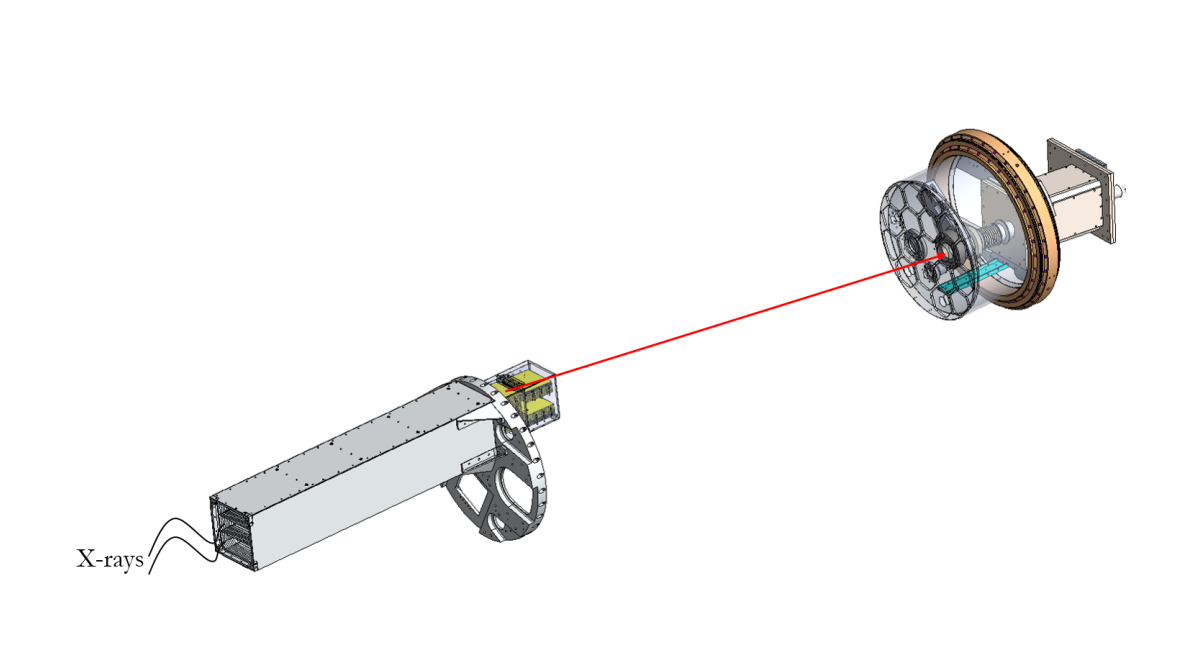}
    \caption{\label{fig:WRXR} Schematic of WRXR spectrometer. X-rays enter the instrument through the collimator on the left before undergoing diffraction in the grating array, and then reflecting onto the detector over 2 meters away.}
   \end{center}
   \end{figure}

The spectrometer design consists of a mechanical collimator, X-ray reflection gratings, a mirror module, and the hybrid CMOS detector camera. Figure \ref{fig:WRXR} shows a schematic diagram of the full instrument light path. Light enters the lightweight wire-grid collimator and is constrained to converge in one dimension. The off-plane reflection gratings then intercept this light before it can reach the focal plane and disperse it to produce spectral lines that would reach the detector with $\sim 190$ mm cross-dispersion extent. The detector at the focal plane is the same H2RG HCD described in Section \ref{sec:hcd} and has an active area of 35 mm $\times$ 35 mm. Because this would result in less than 20\% of a given spectral line falling on the detector area, a mirror module is inserted between the reflection gratings and the camera. This module contains an array of nickel-coated mirrors that reflect light back onto the detector area and minimize photon losses from the gratings. 

The camera package is a custom built $8.5 \times8.5 \times 12.75$ inch enclosure that is mounted at the spectrometer focal plane. Inside, a copper cold finger will cool the aforementioned H2RG and a cryogenic SIDECAR$^{\text{TM}}$ to 130 K and 180 K respectively. Because this H2RG detector does not have an aluminum filter deposited directly onto it, a $450\angstrom$ Ti plus $700\angstrom$ Al filter will be provided by Luxel and installed in front of the detector surface. A $^{55}$Fe calibration source will also be mounted inside the enclosure and angled such that it only produces counts at X-ray line free portions of the detector, enabling valuable in-flight calibration. The detector housing will be isolated from the main instrument vacuum section by a controllable GN2 actuated gate valve, protecting the detector from external factors when not pointed at the science target; an on-board ion pump will maintain vacuum pressure in the enclosure. Outside of the enclosure a custom-built board interacts with the SIDECAR$^{\text{TM}}$ ASIC to provide power, filtering, and data buffering. The H2RG will have 15 V applied to its substrate and run with 32 parallel output channels, providing a frame time of 1.48 s. 

\noindent For a more complete description of the instrument design and full specifications, see Miles et al. (2017) \cite{WRX}.

\section{Summary}

The PSU X-ray detector lab is involved in various activities relating to hybrid CMOS detectors and their continued technological development. New tests with a cryogenic SIDECAR$^{\text{TM}}$ ASIC and H2RG have demonstrated that significantly improved energy resolution and read noise are realized when cooling the SIDECAR$^{\text{TM}}$. Energy resolution of 156 eV at 5.9 keV and read noise of $\sim 6.8$ e$^-$ have been achieved with this H2RG. Small pixel HCDs with 12.5 micron pixel pitch, in-pixel CDS, and CTIA amplifiers have also been characterized, with $\sim 5.5$ e$^-$ read noise measured in the best results. The best overall pixel type for these small pixel HCDs is the pixels that include in-pixel CDS and extra cross-talk reducing shielding. Other efforts in the detector lab include development of a large array Speedster-EXD device, and preparations to launch an HCD on a sounding rocket instrument in April 2018. The latter will be a key test of X-ray HCDs in a space environment, raising X-ray HCDs to NASA TRL 9 and demonstrating their suitability for future space missions. 

\acknowledgments 
We gratefully acknowledge Teledyne Imaging Sensors, particularly Vincent Douence, Mihail Milkov, Steven Chen, and Mark Farris for their very useful troubleshooting assistance and their excellent design work.  This work was supported by NASA grants NNX13AE57G, NNX17AE35G, NNX14AH68G, and NNX17AD87G.



\end{document}